\documentstyle[preprint,aps,prb]{revtex} 
\begin{document}
\draft
\preprint{\today}
\title{Chemical Hardness, Linear Response, and Pseudopotential
Transferability}

\author{A.~Filippetti,$^1$ David Vanderbilt,$^2$ W.~Zhong,$^2$
Yong Cai,$^{2,*}$ and G.B.~Bachelet$^1$}

\address{$^1$Dipartimento di Fisica, Universita' di
Roma La Sapienza, Roma, Italy}

\address{$^2$Department of Physics and Astronomy,
Rutgers University, Piscataway, NJ 08855-0849}

\date{\today}
\maketitle

\begin{abstract}
We propose a systematic method of analyzing pseudopotential
transferability based on linear-response properties of the free atom,
including self-consistent chemical hardness and polarizability.  Our
calculation of hardness extends the approach of Teter\cite{teter} not
only by including self-consistency, but also by generalizing to
non-diagonal hardness matrices, thereby allowing us to test for
transferability to non-spherically symmetric environments.  We apply
the method to study the transferability of norm-conserving
pseudopotentials for a variety of elements in the Periodic Table.  We
find that the self-consistent corrections are frequently significant,
and should not be neglected. We prove that the partial-core correction
improves the pseudopotential hardness of alkali metals considerably.
We propose a quantity to represent the average hardness error and
calculate this quantity for many representative elements as a
function of pseudopotential cutoff radii.  We find that the atomic
polarizabilities are usually well reproduced by the norm-conserving
pseudopotentials.  Our results provide useful guidelines for making
optimal choices in the pseudopotential generation procedure.

\end{abstract}
\narrowtext

\section{Introduction}

Density-functional calculations performed within the framework of
the local-density approximation (LDA) have been demonstrated to
give accurate predictions of many physical properties of
solids.\cite{DFT} The introduction of the pseudopotential
approximation greatly simplifies electronic-structure calculations
by eliminating the need to include atomic core electrons and the
strong potentials responsible for binding them.\cite{PSP}  The
introduction of norm-conserving pseudopotentials by Hamann,
Schl\"uter and Chiang (HSC)\cite{HSC} led to greatly improved
control of transferability errors, and as a result the
pseudopotential approach has since found a wide range of
applications in molecular and solid-state electronic-structure
theory.  Nevertheless, transferability is still an issue in many
calculations, especially when uncomfortably large pseudopotential
cutoff radii have been dictated by the requirements of a modest
plane-wave cutoff in the solid-state calculation, and for atoms
having shallow core shells.

A pseudopotential (PSP) is constructed to replace the all-electron
(AE) atomic potential in such a way that core states are
eliminated.  The most important measure of a pseudopotential is its
transferability, which characterizes the accuracy with which it
mimics the real AE atom in different atomic, ionic, molecular, or
solid-state environments.  Traditionally, the transferability of a
pseudopotential is characterized by two properties:  (i) a
comparison of the scattering properties of the real and pseudo
versions of the free atom or ion, as measured by the logarithmic
derivative of the wavefunction at some diagnostic radius as a
function of energy; and (ii) configuration tests, which check if
the pseudo eigenvalues and total energies track the AE ones for
various excited states of the free atom or ion.  It is important to
note that spherical symmetry is implicit in both of these
approaches, so that neither is capable of giving information about
transferability to anisotropic environments.

Scattering properties are certainly a significant aspect of
transferability: poor scattering properties are indicative of a
poor pseudopotential.  A major contribution of HSC was to show that
the norm-conserving condition automatically implies that not only
the logarithmic derivative, but also its energy derivative, is
guaranteed to be correct for the PSP at a reference energy, thus
insuring that the AE and PSP scattering properties will match
closely over a large energy range.\cite{HSC} Thus norm-conserving
potentials tend to have much better transferability for most
elements.  However, the matching of logarithmic derivatives is not
always sufficient to ensure good transferability.  Some potential
sources of error which will not show up in tests of scattering
properties are (i) electrostatic screening effects, and (ii)
effects of non-linearity of the exchange-correlation energy
(important for many alkali-metal elements).  Errors of the former
type are usually easily eliminated by the choice of a
``conservative'' cutoff radius,\cite{BHS,goedecker} while those
arising from core-valence overlap can largely be corrected by use
of the frozen-core correction.\cite{LFC}  Nevertheless, these
examples illustrate the dangers of focusing on scattering
properties alone.

Configuration tests are certainly useful as a supplementary
criterion, but as mentioned above, they do not control the quality
of the PSP in a nonspherical target environment.  This will
obviously be important for atoms in surface, defective, molecular,
or liquid environments, to name just a few.  Furthermore, it is
difficult to include the configuration tests systematically as part
of the PSP generation procedure.

As LDA calculations are pushed in the direction of high accuracy,
PSP errors become less tolerable, putting tight requirements on
transferability.  On the other hand, as the calculations are pushed
to larger system sizes, the increased computational load requires
that the PSP be as smooth (soft) as possible.  This has led to a
tremendous effort to optimize PSP softness.\cite{vand,rappe,trou}
Unfortunately, transferability and softness are usually
contradictory requirements.  Especially for first-row elements,
attempts to save computational cost frequently result in the use of
a PSP with an uncomfortably large core cutoff radius.  Because of
the above-mentioned electrostatic screening problems, this strategy
may result in a sacrifice of transferability which would be
difficult to detect using the conventional methods.  It is
therefore of great importance to develop improved measures of
transferability which will allow for improved control of PSP errors
in cases like these.  It is especially desirable to develop methods
which work directly at the atomic level, without the need for
painstaking comparisons of pseudo and AE results in molecular and
solid-state environments.

In this paper, we propose to use the linear-response properties of
a reference free atom or ion as a measure of the transferability of
a PSP.  We calculate two kinds of linear-response properties:  a
generalized chemical hardness, and the dipole and higher-moment
susceptibilities.  Chemical hardness measures the derivatives of
electronic eigenvalues with respect to changes of occupation.  It
was recently proposed by Teter\cite{teter} as an important measure
of PSP transferability, partly based on the idea that the chemical
potential and hardness have equal roles in determining electron
charge transfers.  To some extent, the hardness analysis is
redundant with configuration tests (of which it is a kind of
differential version), but it is more systematic and can be
incorporated into PSP generation more easily.\cite{teter}  The
concept of hardness is generalized in this paper to include also
information about the response to non-spherical perturbations.
This is in fact very important, and we shall see that the
rearrangement of charge in a $p$-shell may dominate the error in
the hardness matrix. We further include the self-consistent change
of the wavefunctions in our calculations in order to go beyond the
frozen-wavefunction approximation (FWA) introduced by
Teter.\cite{teter} Finally, the dipole (and quadrupole, etc.)
susceptibility tests measure the ability of the pseudo atom to
imitate the correct AE behavior in a local electric field (or field
gradient, etc.) which may result from an anisotropic solid-state or
molecular environment.

This paper is arranged as follows.  In Sec.\ II we review the
concept of chemical hardness and introduce the formulation for
calculating both spherical and non-spherical hardness.  We also
outline the calculation of the dipole and higher-order
polarizabilities.  In Sec.\ III, we present calculated chemical
hardness matrices and polarizabilities for some representative
atoms chosen from different parts of the Periodic Table.  We
discuss the general trends in the  hardness matrix, and the effect
of the frozen-core correction.  We also proposed a quantity based
on the hardness matrix to characterize the pseudopotential
transferability.  Finally, we conclude in Sec.\ IV.

\section{Theory}

\subsection{Chemical hardness}

Teter defines the chemical hardness matrix $H_{ij}$ within the
LDA as\cite{teter}
\begin{equation}
H_{ij}\>=\>{1\over 2}\>{\partial^2 E[\rho]\over
\partial f_i\partial f_j} \; ,
\label{zero}
\end{equation}
where $E$ is the Janak functional\cite{janak} and $f_i$ is the
occupation number of the $i$th state.  Since the eigenvalue
$\epsilon_i = \partial E / \partial f_i$, we have
\begin{equation}
H_{ij}\>=\> {1\over 2} {\partial \epsilon_i \over \partial f_j} \; .
\label{ehdef1}
\end{equation}
Thus, the hardness matrix measures the first-order change of an
energy eigenvalue resulting from a first-order variation of an
occupation number, while allowing the total number of electrons to
vary.  It consists of two parts: one due to the change of the
screening potential with variation of the occupation number for
fixed wavefunctions, and one arising from relaxation of the
wavefunctions.  Teter made the approximation of omitting the second
term, but both are included here.

Frequently, one is interested only in the case where the occupations
of the states comprising a given shell are kept equal.  For
example, one may consider an excitation in which one transfers
an $s$ electron to the $p$ shell, increasing the occupation of
each $p$ state by 1/3.  This insures retention of spherical
symmetry of the charge density and potential, and is implicit in
all of the analysis which is usually carried out with an atomic
pseudopotential program.  In this case,  different $m$
components remain degenerate and the treatment is simple.\cite{teter}
In real situations (in molecules, at surfaces, etc.),
atoms may have very anisotropic environments, so that non-spherical
changes of electron occupation become important.
This prompts us to consider also changes of occupation which lead
to nonspherical changes of density and screening potential.  We
will use the index $L$ to refer to density or potential changes
having angular character $Y_{LM}(\Omega)$.  Thus, we shall not
restrict ourselves to spherically symmetric ($L=0$) perturbations,
but will consider the general $L\ne0$ case.

For this purpose, it is useful to generalize from the concept of an
occupation number $f_i$ to the concept of an ``occupation matrix''
or ``density matrix'' $f_{ij}$.  This generalization, previously
introduced in other contexts,\cite{vandj,lnv} makes
the hardness analysis more complete.  It is described in Subsec.\ IIC
below.  While $f_{ij}$ is diagonal in the atomic ground state,
or in a basis of energy eigenstates of a perturbed system, it
may be non-diagonal in a more general representation of a
perturbed system such as an atom in a defective environment.

In our calculations, we use first-order density-functional
perturbation theory in the framework of LDA, following the scheme
formulated by Mahan and Subbaswamy.\cite{mahan} In the remainder
of this section, we first give a detailed formulation of the
calculation of the hardness matrix elements associated with
conventional diagonal occupation number changes, and discuss the
extension beyond the frozen-wavefunction approximation (FWA).
Next, the generalization to non-diagonal occupation changes will be
presented.   Finally, we sketch the calculation of the dipole and
higher susceptibilities, which is straightforward after the
machinery needed to calculate the hardness elements has been set
up.

\subsection{Hardness for diagonal occupation}

Within the framework of LDA, application of the
Hellmann-Feynman theorem to Eq. (\ref{ehdef1}) yields
\begin{eqnarray}
 H_{ij} & = & {1\over 2} \langle\psi_i | {\partial \over \partial
f_j} \left[ T + V_{\rm ion} + V_{\rm hxc}  \right]| \psi_i\rangle
\nonumber\\
    & = & {1\over 2} \langle\psi_i
    | {\partial V_{\rm hxc} \over \partial \rho }
      {\partial \rho \over \partial f_j} | \psi_i\rangle \; ,
\end{eqnarray}
where $V_{\rm hxc}$ is the Hartree and exchange-correlation potential.
With atomic quantum numbers, this becomes
\begin{equation}
H_{\vbox{\baselineskip=8pt
    \hbox{$\scriptstyle nlm$}
    \hbox{$\scriptstyle n'l'm'$}}}
= {1 \over 2} \int\int d{\bf r}\>d{\bf r}'\>
n_{nlm}({\bf r})\>w_{\rm hxc}({\bf r},{\bf r}')\>
{\delta \rho({\bf r}')\over \delta f_{n'l'm'}} \; .
\label{zero1}
\end{equation}
Here,
\begin{eqnarray}
 n_{nlm}({\bf r}) & = & R_{nl}^2(r)\>|Y_{lm}(\Omega)|^2 \; ,\\
 w_{\rm hxc} ({\bf r},{\bf r}') & = &
 \frac{\partial V_{\rm hxc} ({\bf r})}{\partial \rho({\bf r}')}
\; , \\
\rho({\bf r}) &=& \sum_{nlm}^{\rm occ}\>f_{nlm}\>n_{nlm}({\bf r}) \;.
\end{eqnarray}

The density variation due to the variation of occupation numbers
consists of two terms:
\begin{equation}
{\delta \rho({\bf r})\over \delta f_{nlm}}=
n_{nlm}({\bf r})+\Delta n_{nlm}({\bf r}) \; .
\label{zero2}
\end{equation}
The first term arises from the explicit dependence of density on
the occupation numbers, while the second involves relaxation of the
wavefunction with changes of occupation.  We will refer to the
neglect of $\Delta n_{nlm}$ as the ``frozen wavefunction
approximation'' (FWA), while the effect of the $\Delta n_{nlm}$
term will be referred to as the ``self-consistency'' (SC)
correction.

The FWA part of the hardness matrix is relatively easy to
calculate, and it has previously been done for almost all atoms in
the Periodic Table.\cite{af}  The SC correction is treated using
density-functional linear response theory, regarding the change of
$f_{nlm}$ as an external perturbation.

We start with the calculation of the FWA hardness, which may be
written
\begin{equation}
H_{\vbox{\baselineskip=8pt
    \hbox{$\scriptstyle nlm$}
    \hbox{$\scriptstyle n'l'm'$}}}
^{\rm FWA}={1\over 2}\int d{\bf r}\>
V_{nlm}({\bf r})\>n_{n'l'm'}({\bf r}) \;,
\label{cagliari}
\end{equation}
where
\begin{equation}
V_{nlm}({\bf r})=\int d{\bf r}'\>w_{\rm hxc}({\bf r},{\bf r}')\>
n_{nlm}({\bf r}') \; .
\label{first}
\end{equation}
The kernel $w_{\rm hxc}$ can be decomposed as
\begin{equation}
w_{\rm hxc}({\bf r},{\bf r}')=\sum_{LM}\>
w_{\rm hxc}^{(L)}(r,r')\>
Y_{LM}^*(\Omega)\>Y_{LM}(\Omega') \; ,
\label{parigi}
\end{equation}
where
\begin{equation}
w_{\rm hxc}^{(L)}(r,r')=
{8\pi\over 2L+1}\>{{r^<}^L \over {{r^>}^{L+1}}}\>
+{\delta V_{xc} \over \delta n(r)}\>{\delta(r-r')\over r^2}
\label{cisterna}
\end{equation}
and $r^<$ and $r^>$ are the smaller and larger of $r$ and $r'$
respectively.  Note that the exchange-correlation term, being
local, is independent of $L$.  The functions $n_{nlm}$ and
$V_{nlm}$ can also be expressed in spherical harmonics as
\begin{equation}
n_{nlm}({\bf r})=\sum_{L=0}^{2l}\>
C^{(L)}(l,m)\>n_{nl}(r)\>
Y_{L,0}(\Omega)   \; ,
\label{antonio}
\end{equation}
\begin{equation}
V_{nlm}({\bf r})=\sum_{L=0}^{2l}\>C^{(L)}(l,m)
\>v_{nl}^{(L)}(r)\>Y_{L,0}(\Omega) \; ,
\label{second}
\end{equation}
where the $C^{(L)}$ are Gaunt coefficients, reflecting the presence
of squared spherical harmonics in $n_{nlm}({\bf r})$.  Here and in
the remainder of this subsection, the sums are over even $L$ only.
Defining
\begin{eqnarray}
h_{nn'll'L}^{\rm FWA} & = & {1\over 2}\>
\int dr\>r^2\>n_{nl}(r)\>v_{n'l'}^{(L)}(r) \nonumber \\
 & = &{1\over 2}\>\int\int dr\>dr'\>r^2\>r'^2\>n_{nl}(r)\>
w_{\rm hxc}^{(L)}(r,r')\>n_{n'l'}(r') \; ,
\label{genny}
\end{eqnarray}
the FWA hardness can be expressed as
\begin{equation}
H_{\vbox{\baselineskip=8pt
    \hbox{$\scriptstyle nlm$}
    \hbox{$\scriptstyle n'l'm'$}}}
={1\over 2}\>\sum_{L=0}^{2l_{\rm min}}
C^{(L)}(l,m)\>
C^{(L)}(l'm')\>
h_{nn'll'L}^{\rm FWA} \; ,
\label{france}
\end{equation}
where $l_{\rm min}$ is the smaller of $l$ and $l'$.

Of course, the presence of $V_{nlm}({\bf r})$ induces a change
$\Delta n_{nlm}({\bf r})$ in the charge density.  The SC correction
to the hardness can be calculated by including this change
self-consistently, treating $V_{nlm}$ as a bare potential
perturbation.  We can write
\begin{equation}
\Delta n_{nlm}({\bf r})=\int d{\bf r}'\>
\chi({\bf r},{\bf r}')
\>V_{nlm}({\bf r}')
\label{actchi}
\end{equation}
where $\chi$ is the linear susceptibility (to be discussed below).
According to a theorem by Eaves and Epstein,\cite{eaves} the
induced charge density for a closed-shell atom has the same angular
character as the perturbing potential.  Within LDA (with
spin-polarization neglected) the ground state is always isotropic
(equal population of each $m$ character of any given shell), so
this theorem applies.  Thus, to linear order we can write
\begin{equation}
\Delta n_{nl}^{(L)}(r)=\int dr'\>r'^2\>
\chi^{(L)}(r,r')
\>v_{nl}^{(L)}(r')\> \,
\label{fourth}
\end{equation}
where $\chi^{(L)}(r,r')$ is the linear susceptibility in radial
coordinates for perturbations of angular character $L$, and
\begin{equation}
\Delta n_{nlm}({\bf r})=
\sum_{L=0}^{2l}\>C^{(L)}(l,m)\>
\Delta n_{nl}^{(L)}(r)\>
Y_{L,0}(\Omega)  \; ,
\label{third}
\end{equation}
in analogy with Eq.\ (\ref{antonio}).  After defining
\begin{equation}
\Delta
h_{nn'll'L}^{\rm SC}=
{1\over 2}\>\int\int dr\>dr'\>r^2\>r'^2\>v_{nl}^{(L)}(r)\>
\chi^{(L)}(r,r')\>v_{n'l'}^{(L)}(r')\>  \; ,
\label{casso}
\end{equation}
the SC correction to the hardness matrix becomes
\begin{eqnarray}
\Delta H_{\vbox{\baselineskip=8pt
    \hbox{$\scriptstyle nlm$}
    \hbox{$\scriptstyle n'l'm'$}}}
 & = & {1\over 2}\>\int\int d{\bf r}\>d{\bf r}'
\>V_{nlm}({\bf r})\>\chi({\bf r},{\bf r}')
\>V_{n'l'm'}({\bf r}') \nonumber\\
 & = & {1\over 2}\sum_{L=0}^{2l_{min}}
C^{(L)}(l,m)\>C^{(L)}(l'm')\>
\Delta h_{nn'll'L}^{\rm SC} \; .
\label{fifth}
\end{eqnarray}

The determination of the linear susceptibility $\chi({\bf r},{\bf
r}')$ follows closely the modified Stern\-heimer approach discussed
by Mahan and Sub\-ba\-swamy.\cite{mahan} In what follows, we
consider the response $\Delta n^{(L)}({\bf r})=\Delta n^{(L)}(r)
Y_{L0}(\Omega)$ to a general perturbation $v^{(L)}({\bf
r})=v^{(L)}(r) Y_{L0}(\Omega)$ of angular character $L$.  (We have
$v=v_{nl}$ and $\Delta n=\Delta n_{nl}$ in the immediate context,
where $nl$ are indices of the state whose occupation is being
varied.  However, the discussion given below is general, and the
indices $nlm$ will henceforth be taken to refer to arbitrary
occupied wavefunctions which respond to the perturbation.) We do
not actually need to calculate $\chi$ itself; instead, it is
sufficient to specify an iterative algorithm for calculating its
action
\begin{equation}
\Delta n^{(L)}(r)=\int dr'\>r'^2\>
\chi^{(L)}(r,r')
\>v^{(L)}(r)\>
\label{chiL}
\end{equation}
upon the arbitrary perturbation.

The procedure is as follows.  For the moment, assume that the
first-order density change $\Delta n^{(L)}(r)$ is known; then the
corresponding change $\Delta v^{(L)}_{\rm scf}(r)$ in the screened
Kohn-Sham potential is given by
\begin{equation}
\Delta v^{(L)}_{\rm scf}(r) = v^{(L)}(r)
+ \int dr'\>r'^2 \, w_{\rm hxc}^{(L)}(r,r') \, \Delta n^{(L)}(r') \;.
\label{edv}
\end{equation}
This induces a first-order change in each Kohn-Sham wavefunction
satisfying
\begin{equation}
\left({\cal H}_0-\epsilon_{nl}\right) \, \Delta \psi_{nlm}=
-\,\left(\Delta V^{(L)}_{\rm scf}-\Delta \epsilon_{nlm} \right) \,
\psi_{nlm} \; ,
\label{ccitua}
\end{equation}
where ${\cal H}_0, \epsilon_{nl}$, and $\psi_{nlm}({\bf r})$ are
the unperturbed Kohn-Sham Hamiltonian, eigenvalue, and
eigenfunction, respectively, and $\Delta \epsilon_{nlm}$ and
$\Delta \psi_{nlm}({\bf r})$ are the corresponding first-order
changes.  Decomposing Eq.\ (\ref{ccitua}) into angular and radial
parts, one finds
\begin{equation}
\Delta \psi_{nlm}({\bf r})  = \sum^{l+L}_{l'=|l-L|} C(L0;lm,l'm)
 \Delta R_{nll'} (r) Y_{l'm} (\Omega)
\label{delpsi}
\end{equation}
where the $C(L0;lm,l'm)$ are Clebsh-Gordon coefficients
and the radial functions $\Delta R_{nll'}(r)$ are the solutions
of the radial part of Eq.~(\ref{ccitua}),
\begin{equation}
\left[-{d^2\over dr^2}+{l'(l'+1)\over r^2}+V_{\rm ion}(r)+
V_{\rm hxc}(r)-
\epsilon_{nl}\right] \Delta R_{nll'}(r)=\left[\Delta \epsilon_{nlm}
-\Delta v_{\rm scf}(r)\right]R_{nl}(r) \; .
\label{erks}
\end{equation}
This inhomogeneous equation is solved numerically on a radial mesh
following Ref.\ \onlinecite{mahan}.  Finally, the change of density
resulting from Eq.\ (\ref{delpsi}) is
\begin{equation}
\Delta n^{(L)}(r)\>=\sum_{n,l,l'}
\frac{f_{nl}}{2(2l+1)}
\Delta R_{nll'}^{(L)}(r)\>R_{nl}(r)D(ll',L)\>P_L(\Omega) \; ,
\label{edna}
\end{equation}
where $f_{nl}=\sum_m f_{nlm}$, and the geometric coefficients
$D(ll',L)$ are given on p.\ 55 of
Ref.\onlinecite{mahan}.  Iterative solution of Eqs.\ (\ref{edv}),
(\ref{erks}), and (\ref{edna}) thus gives the self-consistently
screened density change $\Delta n^{(L)}(r)$ resulting from the
perturbation $v^{(L)}(r)$ in Eq.\ (\ref{chiL}).

\subsection{Non-diagonal hardness}

If one wants to express {\it any} atomic charge density in terms of
(both filled and empty) Kohn-Sham orbitals of a reference
ground-state atom, then in general the occupation numbers turn out
to be non-diagonal, so that the total charge density should be
expressed as
\begin{equation}
n({\bf r})=\sum_{ij}\>f_{ij}\>\psi_i^*({\bf r})\psi_j({\bf r})
\label{latina}
\end{equation}
where $f_{ij}=f_i\>\delta_{ij}$ only in the ground state.
The most general result of a perturbation is thus obtained by
allowing for non-diagonal terms $f_{ij}$ to exist,\cite{vandj,lnv}
and by taking this into account we arrive at a more general form
of the hardness matrix.  This provides a natural way to test the
transferability of a PSP to nonspherical environments.

Instead of working in the explicit atomic representation
$f_{ij}=f_{nlm,n'l'm'}$, we find it convenient to work in a
representation $(nn'll'LM)$ in which $LM$ are labels of total
angular momentum, and $L= |l-l'|, |l-l'|+2,...,l+l'$ following the
usual angular-momentum addition rules.  Introducing also the
condensed notation $\alpha=nn'll'$, we thus have
\begin{equation}
n({\bf r})=\sum_{\alpha LM}\>f_{\alpha LM}\>n_{\alpha}(r)\>
Y_{LM}(\Omega)
\label{roma}
\end{equation}
where
\begin{equation}
f_{\alpha LM}=\sum_{mm'} C(LM;lm,l'm') f_{nlm,n'l'm'}
\label{ftof}
\end{equation}
with $C(LM;lm,l'm')$ again being the Clebsh-Gordon coefficients, and
$n_{\alpha}(r)=R_{nl}(r)R_{n'l'}(r)$.
In this notation, it is natural to introduce the generalized
Kohn-Sham eigenvalues
\begin{equation}
\epsilon_{\alpha LM}={\partial E\over \partial f_{\alpha LM}}
\label{milano}
\end{equation}
and the generalized hardness matrix
\begin{equation}
H_{\alpha \beta LL'MM'}={1\over 2}{\partial^2 E\over
\partial f_{\alpha LM}\>\partial f_{\beta L'M'}} \; .
\label{firenze}
\end{equation}

Since our reference unperturbed configuration is spherically
symmetric, $\epsilon_{\alpha LM}$ is only nonzero for $L=M=0$ and
$n=n'$, $l=l'$.
Spherical symmetry also implies that the hardness matrix reduces to
\begin{equation}
H_{\alpha \beta LL'MM'}=H_{\alpha \beta L} \, \delta_{LL'} \,
\delta_{M,-M'}  \; .
\label{spherchi}
\end{equation}
The FWA hardness becomes just
\begin{equation}
H_{\alpha\beta L}^{\rm FWA}={1\over 2}\>\int\int dr\>dr'\>r^2\>r'^2\>
n_{\alpha}(r)\>w_{\rm hxc}^{(L)}(r,r')\>n_{\beta}(r') \; ,
\label{londra}
\end{equation}
where the quantity $w_{\rm hxc}^{(L)}$ is as defined in Eq.\
(\ref{cisterna}) (now
generalized to odd as well as even $L$).  The calculation of the SC
correction proceeds along similar lines as for the diagonal case.
The perturbing potential is generalized to
\begin{eqnarray}
V_{\alpha LM}({\bf r}) & = & \int dr'\>r'^2\>w_{\rm hxc}^{(L)}(r,r')
\>n_{\alpha}(r')\>Y_{LM}(\Omega)  \nonumber\\
& = & v_{\alpha}^{(L)}\>(r) Y_{LM}(\Omega)
\label{madrid}
\end{eqnarray}
and the self-consistent correction is given by
\begin{equation}
\Delta H_{\alpha\beta L}^{\rm SC}={1\over 2}\>
\int\int dr\>dr'\>r^2\>r'^2\>v_{\alpha}^{(L)}(r)\>
\chi^{(L)}(r,r')\>v_{\beta}^{(L)}(r') \; .
\label{torino}
\end{equation}

In the special case $\alpha=nnll$ and $\beta=n'n'l'l'$, the
quantities $H_{\alpha \beta L}^{\rm FWA}$ and $H_{\alpha \beta
L}^{\rm SC}$ reduce to $h_{nn'll'L}^{\rm FWA}$ and
$h_{nn'll'L}^{\rm SC}$, respectively, as defined in the previous
subsection.  Thus, all of the diagonal hardness matrix elements are
contained as special cases of the generalized non-diagonal ones
introduced here.  Moreover, the original diagonal formulation only
covers variations of the screening potential or density of even $L$
(monopole, quadrupole, etc.), whereas the generalized formulation
is capable of treating variations of any angular character.  In
fact, perturbations of dipole ($L=1$) character are likely to be
the most important nonspherical perturbations in many molecular and
solid-state environments, especially at surfaces and other defects
where inversion symmetry is lacking.  Therefore, in what follows we
will concentrate on comparisons of the AE and PSP non-diagonal
hardness elements $H_{\alpha \beta L}$, with special emphasis on
the $L=0$ and $L=1$ cases.

\subsection{Polarizability}

The polarizability of an atom measures its response to an external
electric field.  The dipole ($L=1$), quadrupole ($L=2$), and higher
($L>2$) polarizabilities are defined as the derivatives of the
$L$'th induced charge moment with respect to an electrostatic
potential of form $r^L Y_{L0}(\Omega)$.  For good transferability,
it is important that the pseudo-atom have polarizabilities similar
to those of the all-electron atom.  We can expect the lower-moment
polarizabilities to be more important, so we focus on the dipole
and quadrupole susceptibility in what follows.  Tests of
the non-selfconsistent polarizability of HSC pseudopotentials have
previously been performed for a large number of closed-shells atoms
and ions by Bachelet {\it et al.}\cite{bach1}  Here, we extend the
tests to other atoms and also include the SC correction.

The formulation and calculation of the polarizability is
straightforward\cite{mahan,mahan1} using the machinery developed
in the previous subsections.  The perturbing potential is
taken to have the form
\begin{equation}
V_L({\bf r})=r^L Y_{L0} (\Omega) \; ,
\label{caserta}
\end{equation}
and the linearly induced density change is
\begin{equation}
\Delta n^{(L)}(r)=\int dr'\>r'^{L+2}\>
\chi^{(L)}(r,r')  \;.
\end{equation}
This is evaluated using the same iterative procedure given
previously in Eqs.\ (\ref{edv}-\ref{edna}).
The polarizability in angular channel $L$ is then
\begin{equation}
p_L=-\frac{8\pi}{2L+1} \int dr\> r^{L+2} \> \Delta n^{(L)}(r) \; ,
\label{trento}
\end{equation}
with $p_1$ and $p_2$ being the dipole and quadrupole
susceptibility, respectively.

\section{Results}

In this part, we present our calculated hardness matrix elements
for a set of representative atoms.  We begin with all-electron
atoms, surveying the characteristics of the hardness matrix and the
basic trends as a function of position in the Periodic Table.  We
concentrate first on argon, making the comparison between the AE
and PSP hardness matrix, discussing sources of error, and
introducing our format for presenting results in a systematic
fashion.  Next, we discuss trends as one goes across the Periodic
Table, and evaluate the importance of the Louie-Froyen-Cohen
(LFC)\cite{LFC} semi-core correction and the effects of varying the
core radius.  We will then propose a method to extract the most
important information from the large number of hardness elements.
Finally, the calculated AE and PSP polarizabilities will be
presented at the end of this section.

The results reported here are restricted to potentials of the HSC
type.  We believe that these results can be taken as indicative for
the whole class of norm-conserving HSC-like
PSPs,\cite{HSC,BHS,trou,vand,rappe,kerker} provided comparable
cutoff radii are chosen.  For Kleinman-Bylander,\cite{kb}
ultrasoft,\cite{ultrasoft} or other approaches to PSP construction
which deviate significantly from the original HSC method, addition
investigation may be appropriate.

\subsection{Hardness matrix for all-electron atoms}

Before studying the transferability of PSPs with reference to the
hardness matrix elements $H_{ij}$, we first need to understand the
behavior of $H_{ij}$ for all-electron atoms in order to obtain some
physical intuition.  In Table \ref{boia}, we show some matrix
elements of the generalized hardness for three characteristic AE
atoms in the Periodic Table.  Silicon and argon are taken in the
ground-state neutral configuration, while for sodium we choose an
ionized valence configuration $3s^{0.5}3p^{0.25}$ in order to
ensure that the $p$ electron is bound.  The total self-consistent
hardness
$H^{\rm total}=H^{\rm FWA} + H^{\rm SC}$ is broken down into
frozen-wavefunction-approximation and self-consistent-correction
pieces as discussed in the previous section.  For some elements,
the FWA contribution is further separated into Hartree (h) and
exchange-correlation (xc) contributions, $H^{\rm FWA}=  H^{\rm h} +
H^{\rm xc}$.  The matrix elements can be identified by quantum
numbers $\alpha=n_1n_1'l_1l_1'$, $\beta=n_2n_2'l_2l_2'$, and $L$.
Since the principle quantum numbers $n$ are obvious in most cases,
we will usually omit them and write the indices simply as
$l_1l_1',l_2l_2';L$.

{}From Table I, we see that $H^{\rm FWA}$ typically dominates,
and the self-consistent correction $H^{\rm SC}$ is only about
$20-30$\% of $H^{\rm FWA}$.  Nevertheless, this is clearly large
enough that complete neglect of $H^{\rm SC}$ is unjustified.  It is
also obvious from the table that $H^{\rm FWA}$ are mostly positive,
while the $H^{\rm SC}$ are negative.  For the diagonal elements
(e.g., excluding $\{ sp,sp;L=0 \}$), this can be understood as
follows.  From Eq.\ (\ref{londra}), the Hartree part of $H^{\rm
FWA}$ can be seen to have the form of the Coulomb self-energy of a
particular charge distribution $n_\alpha(r)Y_{LM}(\Omega)$, which
must be positive.  On the other hand, the exchange-correlation
contribution is negative because $\delta V_{xc}/\delta n$ in
Eq.\ (\ref{cisterna}) is negative.  Typically, the Hartree term
dominates and $H^{\rm FWA}$ is positive.  Regarding $H^{\rm SC}$,
note that Eq.\ (\ref{torino}) can be interpreted as evaluating the
second-order energy change of the system when the external
potential of Eq.\ (\ref{madrid}) is applied; the total energy must
go down when the wavefunctions relax, so that $H^{\rm SC}$ must be
negative.

Going from Na to Ar, we find that $H^{\rm total}$ increases
strongly, primarily because of the Hartree contribution $H^{\rm h}$
to $H^{\rm FWA}$.  This is in agreement with our intuition, since
the wavefunctions become much more localized as one moves from left
to right across the Periodic Table, causing $H^{\rm h}$ to increase
sharply.  In this sense, we can refer to atoms (like Ar) on the
right side of the Periodic Table as ``strongly electrostatic
atoms'' or ``hard atoms,'' while those on the left side (like K)
can be termed ``soft.''

Our results also indicate that the matrix elements of $H^{\rm
total}$ for $L=0$ are significantly larger than those for $L>0$.
This implies that the diagonal hardness elements defined in
Eqs.\ (\ref{france}) and (\ref{fifth}) are dominated by the
spherically symmetric part of the response.  A closer inspection
reveals that $H^{\rm h}$ decreases strongly with increasing $L$,
while $H^{\rm xc}$ is smaller and, being a local operator,
independent of $L$.  This gives rise to an overall small $H^{\rm
total}$ for high $L$.  In a few cases where the valence
wavefunction was very weakly bound and delocalized, we have found
that $H^{\rm xc}$ can even be larger in magnitude than $H^{\rm h}$,
resulting in a small but {\it negative} $H^{\rm total}$.  We regard
this as an unphysical result which reflects the overestimate of
exchange-correlation effects by the LDA in the low-density tail of
the atom.

\subsection{Argon: AE and PSP hardness}

Having gained some understanding of the AE hardness matrix, we turn
now to a comparison of the PSP hardness elements with the
corresponding AE ones.  We begin with argon.  In Table
\ref{t_argon} we list values for some important hardness matrix
elements calculated for the AE and PSP $Ar^+$ ion in configuration
$s^{1.2}p^{5.7}d^{0.1}$.   The HSC PSP was generated in this
configuration, using a core radius $r_c \simeq 1$ a.u.  We also show
the relative errors for another valence configuration,
$s^1p^{5.5}d^{0.5}$.

Comparing the AE and PSP results, we find very good agreement for
the $L=0$ matrix elements.  Norm conservation imposes the
constraint that the $L=0$ component of the electrostatic potential
in the PSP case should match the AE one outside the core
region.  Consequently, for small core radii
the differences between AE and PSP values of $H^{\rm h}$ are
essentially confined to $L>0$ moments.  Actually, we find that the
errors in $H^{\rm h}$ are relatively insensitive to a modest
increase of core radius for hard atoms like Ar.
Because of this constraint,
while the Hartree contribution is large in magnitude, it may only
incur a small error in hardness.
For Ar, the $H^{\rm
xc}$ are quite small and their contribution to the error is not
very significant.  So, for Ar, the agreement between AE and PSP
hardness elements is excellent in the spherically-symmetric ($L=0$)
channel.  The relative errors for $L>0$ channels are larger ($\sim
20\%$ for $L=1$ and $\sim4\%$ for $L=2$).  However, since their
absolute magnitudes are small, they are not as important.  As a
result, the overall agreement between AE and PSP hardness is
very good.  We also checked that changing the testing configuration
affects the results only very slightly.

In order to avoid presenting numerous cumbersome tables, we have
converted the information into the form of a bar chart as shown in
Fig.\ \ref{f_ar}.  The heights of the bars represent the values of
the corresponding hardness matrix elements.  The columns indicate
different contributions to a given hardness matrix element, whose
indices are labeled by the row.  The results for all-electron and
pseudo atoms are placed side-by-side to facilitate comparison;
hollow bars indicate AE results while dashed bars represent PSP
results.  The values for $L > 0$ elements, being small, are
magnified in the diagram.  We believe the discussion is easier to
follow by viewing such diagrams, so all subsequent hardness results
will be presented in this way.

\subsection{From Ar to K}

Argon is a rare-gas element. To explore the general trends of the
hardness matrix along the Periodic Table, we further calculated
the hardness matrix for atoms having a wide range of properties.

We start with Si.  To study the effect of core radius $r_c$ on the
quality of PSP generated, we plot in Fig.\ \ref{f_si} the results
for both small and large $r_c$ indicated by dashed and shaded bars
respectively (values for $r_c$ are indicated in the captions). The
AE results are still plotted with hollow bars.  These results are
for Si$^+$ in configuration $s^2p^1d^0$.  For small $r_c$, the PSP
hardness elements for Si agree with the AE results very well (the
biggest errors occur for $L=1$ elements as for Ar).  With the
exception of $\{ dd,dd; L=0 \}$ elements, a worsening of the
agreement in total hardness is evident in the $L=0$ channel for
larger $r_c$.  (The $d$ wavefunction is nodeless and very
delocalized, and thus insensitive to changes in core radius.)

One expects that increasing $r_c$ should always make the PSP less
transferable, but the sensitivity can be different for different
elements.  In Fig.\ \ref{f_o}, we show a similar diagram for
oxygen, again using hashed and shaded bars to represent
hardness elements for a PSP with small
and large $r_c$.  We find
that the effect of increasing $r_c$ is more dramatic for O.  A more
complete picture of the effect of $r_c$ will be presented later.

We next focus on some cases to characterize the role of the LFC
correction.\cite{LFC} In the following figures, we use the shaded
bars to represent PSP results with such LFC correction.  The
results we presented are for K$^{+0.25}$ in
$s^{0.25}p^{0.25}d^{0.25}$ (Fig.\ \ref{f_k}), Ti$^{+0.75}$ in
configuration $s^1p^{0.25}d^2$ (Fig.\ \ref{f_ti}), C$^{+0.4}$ in
$s^{1.9}p^{1.4}d^{0.3}$ (Fig.\ \ref{f_c}), and Ga$^{+0.5}$ in
$s^2p^{0.5}d^0$ (Fig.\ \ref{f_ga}).  Starting with K we find as
expected that the LFC greatly reduces the error due to the $H^{\rm
xc}$ contribution to the hardness matrix elements even for the
$L=0$ channel. The LFC-corrected hardness matrices are in good
agreement with the AE results.  For Ti, the LFC successfully
corrected the noticeable mismatch between AE and no-LFC results for
$H^{\rm xc}$ contributions to the $\{ dd,dd;L=0 \}$ and $\{
dd,dd;L=2 \}$ elements.  (The no-LFC total hardness in the $\{
dd,dd;L=0 \}$ happens to match the AE one rather closely, but only
because of a fortuitous cancellation of errors.) For C and Ga, the
effect is small except for some high-$L$ channels (i.e. $\{
pp,pp;L=2 \}$ elements), and we regard the LFC as less necessary
for these elements.

The hardness results obviously depend on details of the PSP
construction, but we note the following general trends.  In going
from left to right across a row of the Periodic Table, an
increasing atomic number tends to localize the core density closer
to the nucleus and to reduce the size of the core.  Consequently,
the overlap between core and valence charge densities, which is the
source of the nonlinearity in the exchange-correlation potential,
gets smaller and smaller.  Such an overlap, when significant, is
largely responsible for the errors in the self-consistent
contribution to hardness as well, i.e., for a poor description of
the rearrangement of the pseudo wavefunctions.  Provided that one
takes small core radii to minimize errors in the Hartree
contributions, the degree of core-valence overlap almost entirely
determines the PSP transferability.  Rare gas atoms thus have
maximum transferability without LFC, while alkali atoms need the
LFC correction the most.

Moving along the {\it columns} of the Periodic Table, the trends
are naturally much weaker.  The net positive charge seen by the
valence electrons doesn't change, and the localization of the core
and valence charge densities changes only marginally.  However, a
slight increase of overlap occurs, along with a corresponding loss
of transferability of non-LFC PSP, as one goes down the columns.

{}From our hardness matrix results, we can investigate whether there
is a systematic way to improve the PSP transferability.  The
Hartree contribution can be improved by imposing additional
conditions on the pseudo wavefunctions, e.g., matching of the
valence electrostatic potentials for higher-order multiple
moments.\cite{shirley} However, it is not clear whether the gain in
transferability would justify the drawbacks of imposing additional
constraints.  As regards the exchange-correlation terms, while some
other approaches have been tried,\cite{byla} the LFC correction
seems to be the simplest and most efficient method.

\subsection{Average hardness errors}

We have shown that it is useful to characterize the transferability
of a PSP in terms of its hardness matrix.  However, there is so
much information contained in the numerous matrix elements of the
hardness matrix that it becomes difficult to decide whether a
particular PSP shows ``good'' or ``poor'' transferability.  Thus,
it is desirable to define a single quantity that can be used to
represent approximately the overall transferability of the PSP.
There is certainly no unique way to do this, since the importance
of different matrix elements depends on the target application.
Nevertheless, we propose one such definition, which at least can be
used as a starting point.

We define an average hardness error $X$ as follows:
\begin{equation}
X^2 \>=\>\sum_{\alpha\beta L}\>w_{\alpha\beta L}\>
(\Delta H_{\alpha\beta L})^2 \; .
\label{muzio}
\end{equation}
Here $\Delta H$ is the difference between total AE and PSP hardness
matrix elements, and $w_{\alpha\beta L}$ is a weight to be defined
shortly.  Thus, $X$ is just a weighted RMS average of the errors in
the hardness matrix elements.

To fix the weights $w_{\alpha\beta L}$, we have adopted the
following philosophy.  We want $X$ to represent an average
total-energy error which would occur as the PSP atom is transferred
to an ensemble of target environments, where the distribution of
target environments is characterized by specifying the average
occupation $N_l$ and the typical fluctuation in occupation
$\eta_l$, for each electron shell.  In the spirit of the hardness
approach, we can estimate the change in total energy of the atom as
it is inserted into a given environment as
\begin{equation}
\Delta E \> = \> \sum_{\alpha\beta LM} H_{\alpha\beta L}
\> \delta f_{\alpha LM} \> \delta f_{\beta LM} \;.
\label{ensemble}
\end{equation}
If each contribution were statistically independent, one would have
\begin{equation}
(\Delta E)^2 \> \simeq \> (2L+1) \> \sum_{\alpha\beta L}
H_{\alpha\beta L}^2
\> \delta f_{\alpha LM}^2 \> \delta f_{\beta LM}^2 \;.
\label{ensemble2}
\end{equation}
Making the additional rough approximation that
$\delta f_{\alpha LM}^2\propto \eta_{l_\alpha} \eta_{l'_\alpha}$
and
$\delta f_{\beta LM}^2\propto \eta_{l_\beta} \eta_{l'_\beta}$
(where $\alpha=n_\alpha n'_\alpha l_\alpha l'_\alpha$ and the $n$
subscripts are suppressed) and replacing $(H_{\alpha\beta L})^2$ by
the AE vs.\ PSP error $(\Delta H_{\alpha\beta L})^2$, we arrive at
the right-hand side of Eq.\ (\ref{muzio}) with
\begin{equation}
w_{\alpha \beta L}\>=\>(2L+1)
\> \eta_{l_\alpha} \> \eta_{l'_\alpha}
\> \eta_{l_\beta} \> \eta_{l'_\beta} \;,
\label{weights}
\end{equation}
where it remains to fix the occupation fluctuation $\eta_l$.
To make things simple, we assume that the fluctuation $\eta_l$ is
a function only of the average occupation $N_l$ and the maximum
occupation $2(2l+1)$ of the shell.  We choose the form
\begin{equation}
\eta_l \> = \> (2l+1) \> \sqrt{2f_l(1-f_l)}
\label{etas}
\end{equation}
where $f_l$ is the fractional occupancy, $f_l=N_l/2(2l+1)$.
The first term makes the fluctuation proportional to the number
of electrons which could be accommodated in the shell, and the
second forces the fluctuation to zero for either a completely
filled or a completely empty shell in a manner which respects
electron-hole symmetry.

Thus, we have defined $X$ through Eqs.\ (\ref{muzio}),
(\ref{weights}) and (\ref{etas}), in such a way that it depends
only on a specification of the reference electronic configuration
(the $N_l$ values).  Our definition puts no weight on completely
filled or completely empty shells, and heavily weights
partially-filled shells.

We would be the first to admit that the choices above are largely
arbitrary, but we believe they are reasonable ones, and we proceed
to use this measure to study the effect of variations in core
radius upon PSP transferability. In Figs.\ \ref{f_av1} and
\ref{f_av2}, we show the calculated average hardness error $X$ for
a set of six elements as a function of $r_c = (r_{cs} + r_{cp})/2$.
Here, $r_{cp} - r_{cs}$ is kept constant.  We do not change $r_{cd}$,
since it does not affect $X$ appreciably for the atoms studied
here (with no $d$ electrons inside the core).
The hardness error $X$ (and the hardness itself) is much greater
for first-row elements C and O, so they are plotted on a different
scale.  It can be seen that the behavior of $X$ differs
considerably between elements.  For K, $X$ is very insensitive to
$r_c$, and there is thus wide flexibility in the choice of an
appropriate $r_c$.  For O, C, and Ga, $X$ increases in an
approximately linear fashion as $r_c$ is increased from 1.5 to 3.5
au, while $X$ increases more rapidly for Si and Ar as $r_c$ is
increased.

\subsection{Polarizability}

In Table \ref{sondrio}, we show our calculated
dipole ($L=1$) and quadrupole ($L=2$) polarizability
for some AE and PSP atoms.
Our AE results for rare-gas and closed-shell ions
agree very well with previous calculations.\cite{mahan,zang,stott}
The results for the dipole polarizability are also in
good agreement with experiment. (Experimental values for
higher-moment polarizability do not appear to be available.)
For example, for the K$^+$ ion, we find
the dipole susceptibility to be 5.74 a.u., compared to
5.86 a.u. and 5.47 a.u. from previous theory\cite{mahan} and
experiment\cite{freeman}, respectively.
Note that most of the results reported below are for open-shell
atoms or ions.  It should be emphasized that in these cases, our
results are a theoretical fiction in that we assume symmetrized
occupations (e.g., $s^2p_x^{2/3}p_y^{2/3}p_z^{2/3}$ for C) which
have little relation to the real atomic ground state.
Nevertheless, we believe it is meaningful to compare AE and PSP
polarizability calculated in this way as a means of testing the
transferability of a PSP.  We tend to prefer tests on ionized
configurations (e.g., Si$^+$ instead of Si) because we have found
that shallow orbitals in neutral open-shell atoms sometimes give
such enormous contributions to the polarizability that comparison
becomes difficult.  All PSPs are built choosing small core radii
(e.g., $r_{cs}=1.1$ a.u. for Si$^+$), and for K the LFC correction
was used.

We first consider the all-electron results.  To test the effect of
self-consistent screening, we report both frozen-wavefunction
approximation (FWA) and self-consistent (SC) polarizabilities.  All
pseudopotentials are built choosing small core radii, and for
potassium, the LFC correction is used.  In all cases, the core
polarizability is included in the all-electron value.  As a
example, we show the core contribution of $K^+$ to its all-electron
polarizability.

Looking at Table \ref{sondrio}, it is evident that only the dipole
polarizability is strongly affected by self-consistent screening.
The screening reduces the dipole polarizability by around 40\%,
while the quadrupole susceptibility is typically reduced by only
about 3\%.  Exceptions to this pattern occur for some highly
polarizable atoms like potassium, for which the screening
correction is still sizeable in the quadrupole channel.

Three factors contribute to the difference between PSP and AE
polarizability: (i) the core contribution; (ii) the difference
between unperturbed pseudo and AE wavefunctions inside the core
region; and (iii) differences in the first-order changes in the
valence wavefunctions.  Regarding (iii), the wavefunction
changes are determined in part by admixture of angular-momentum
components higher than those which are present in the unperturbed
reference configuration. For these components (typically $l\ge3$),
no norm-conservation or tail-matching conditions were imposed.
Because the PSP usually contains no nonlocal projectors for
large $l$, these wavefunctions feel only the local potential,
which is usually set in a very arbitrary manner.
This appears to be the most significant source of error in PSP
polarizability.  For example, we have calculated the PSP
polarizability of Ar both with and without an $f$ component in
the nonlocal projector.  We find that the calculated values for
both the dipole and quadrupole polarizabilities are about 15\% too
small when the $f$ component of the projector is omitted.  This
effect was already noted in Ref.\onlinecite{bach1}, where many
other examples can be found.  All the results in Table
\ref{sondrio} are obtained using a PSP with a complete projector
(up to $l=3$).

The results in the Table indicate that while the screening
correction generally improves the agreement between AE and PSP for
the dipole susceptibility, the error in the quadrupole
susceptibility is almost unaffected.  Generally, we find a very
good agreement between the self-consistent AE and PSP results.
However, it should be noted that all results shown in the Table
were obtained using a PSP generated from the same configuration for
which the polarizability calculation was made.
If we change the PSP reference configuration considerably, larger
changes in the calculated pseudo-polarizability may occur.

\section{Conclusion}

We present a systematic method for characterizing the
transferability of pseudopotentials using their linear-response
properties, specifically their generalized chemical hardness and
dipole and quadrupole polarizabilities.  The hardness measures the
ability of the PSP to resemble the all-electron atom in
different atomic environments, including non-spherical ones, while
the polarizability reflect the response of the PSP atom to
external fields.  When used together with conventional criteria such
as norm-conservation and matching of eigenvalues  and logarithmic
derivatives, this approach allows a rather complete
characterization of PSP transferability.

We have applied the method to study the behavior of
Hamann-Schl\"uter-Chiang pseudopotentials for many atoms in the
Periodic Table.  As expected, the calculated hardness matrix
indicates that the transferability deteriorates as the core radius
is increased.  For some elements with relatively delocalized cores,
we find strong evidence for the importance of including the
Louie-Froyen-Cohen semi-core correction.  We propose a method for
reducing the large amount of information contained in the hardness
matrix to a single number.  We suggest that this quantity be
monitored or included in the fitting procedure when generating
pseudopotentials, in order to achieve the desired properties (e.g.,
optimal smoothness) without sacrificing transferability.

\acknowledgments

DV and WZ gratefully acknowledge support from NSF Grant
DMR-91-15342 and ONR Grant N00014-91-J-1184.  AF and GBB
acknowledge support from the Progetto Finalizzato Sistemi
Informatici e Calcolo Parallelo of the Italian National
Research Council, CNR Grants 89.0006.69 and 89.00051.69.
AF also thanks R.~W.~Nunes for his kind help and the Department
of Physics and Astronomy of Rutgers University for its warm
hospitality during part of this project.


\begin{table}
\caption{Calculated all-electron hardness matrix elements of three
representative atoms Na, Si, and Ar.  Both frozen-wavefunction
approximation (FWA) and self-consistent (SC) contributions to the
total hardness elements are listed.  For some matrix elements, the
FWA hardness is broken down into Hartree (h) and
exchange-correlation (xc) contributions.
\label{boia}}
\begin{tabular}{lrrr}
& Na($3s^{0.5} 3p^{0.25}$)  &  Si($3s^2 3p^2$)  &  Ar($3s^2 3p^6$)  \\
    \hline
$ss,ss; L=0$ \\
 FWA   &  0.1906 &0.3923  & 0.6142  \\
 SC  & -0.0166 &-0.0931 & -0.1684  \\
 total & 0.1741  &0.2992  & 0.4457  \\
    \hline
$ss,pp; L=0$ \\
 FWA   & 0.1562  &0.3437  & 0.5668   \\
 SC  & -0.0144 &-0.0737 & -0.1444 \\
 total & 0.1417  &0.2700  &0.4224    \\
    \hline
$pp,pp; L=0$ \\
 FWA(h)& 0.1625  &0.3260  &0.5465   \\
 FWA(xc)&-0.0374  &-0.0197 &-0.0197  \\
 FWA   & 0.1251  &0.3063  &0.5268  \\
 SC  & -0.0131 &-0.0586 &-0.1240  \\
 total & 0.1120  &0.2477  & 0.4028   \\
    \hline
$sp,sp; L=1$ \\
 FWA(h)&0.0367   &0.0766  &0.1275   \\
 FWA(xc)&-0.0319  &-0.0203 &-0.0212  \\
 FWA   & 0.0048  &0.0563  &0.1063   \\
 SC  &-0.0014  &-0.0306 &-0.0560 \\
 total & 0.0033  &0.0257  & 0.0504 \\
    \hline
$pp,pp; L=2$ \\
 FWA(h)&0.0164   &0.0321  &0.0536 \\
 FWA(xc)&-0.0374  &-0.0197 &-0.0197 \\
 FWA   & -0.0210 &0.0124  &0.0339 \\
 SC  & -0.0031 &-0.0018 &-0.0068 \\
 total &-0.0242  &0.0106  &0.0272 \\
\end{tabular}
\end{table}

\begin{table}
\caption{All-electron (AE) and HSC pseudopotential (PSP) hardness
matrix elements for Ar.  Two different electronic configurations
are considered.  The error is the percentage difference between
PSP and AE hardness.  Total hardness (total) is
decomposed into frozen-wavefunction approximation (FWA) and
self-consistent (SC) contributions, while the FWA is further
decomposed into Hartree (h) and exchange-correlation (xc)
contributions.
\label{t_argon}}
\begin{tabular}{lrrrr}
    & \multicolumn{3}{c}{$s^{1.2}p^{5.7}d^{0.1}$}
& {$s^1p^{5.5}d^{0.5}$}\\  \hline
    &  AE (Ry) & PSP (Ry)  & error (\%) & error (\%) \\ \hline

$ss,ss; L=0$ \\
 FWA(h)   &0.6588   &0.6562 &0.39  &0.41\\
 FWA(xc)  &-0.0273 &-0.0308 &12.98  &13.09\\
 FWA      & 0.6315 &0.6254  & 0.97 &1.01  \\
 SC     & -0.1421 &-0.1367& 3.80 &3.26 \\
 total    & 0.4894 &0.4887  &0.15  &0.07 \\
    \hline
$pp,pp; L=0$ \\
 FWA(h)    &0.5876   &0.5845  &0.53 &0.54  \\
 FWA(xc)   &-0.0230  &-0.0254 &10.24 &10.51  \\
 FWA       & 0.5646  &0.5591  &0.97 &1.00  \\
 SC      & -0.1139 &-0.1092 &4.17 &3.44  \\
 total     & 0.4507  &0.4500  &0.16 &0.11 \\
    \hline
$dd,dd; L=0$ \\
 FWA(h)    &0.3237   &0.3234  &0.10 &0.12  \\
 FWA(xc)   & -0.0341 &-0.0344 &0.81 &1.81  \\
 FWA       &0.2897   &0.2890  &0.24 &0.29  \\
 SC      & -0.0380 &-0.0375 &1.24 &1.37  \\
 total     & 0.2517  &0.2515  &0.09 &0.05   \\
    \hline
$sp,sp; L=1$ \\
 FWA(h)    &0.1365   &0.1417  &3.84 &3.91  \\
 FWA(xc)   &-0.0244  &-0.0274 &12.12 &12.25   \\
 FWA       &0.1120   &0.1144  &2.13 &2.05   \\
 SC      &-0.0475  &-0.0342 &27.91 &27.45  \\
 total     &0.0645   &0.0801  &24.25 &21.17  \\
    \hline
$sp,pd; L=1$ \\
 FWA(h)    &0.0935   &0.0987  &5.56 &5.87  \\
 FWA(xc)   &-0.0160  &-0.0178 &11.12 &12.03  \\
 FWA       &0.0775   &0.0809  &4.41 &4.58  \\
 SC      &-0.0331  &-0.0254 &23.23 &22.75  \\
 total     &0.0443   &0.0555  &25.30 &22.54  \\
    \hline
$pp,pp; L=2$ \\
 FWA(h)    &0.0586   &0.0606  &3.33 &3.31\\
 FWA(xc)   &-0.0230  &-0.0254 &10.24 &10.51 \\
 FWA       &0.0356   &0.0352  &1.13 &1.43\\
 SC      &-0.0055  &-0.0037 &32.48 &31.92\\
 total     &0.0301   &0.0315  &4.60 &3.87\\
    \hline
$dd,dd; L=2$ \\
 FWA(h)    &0.0283   &0.0286  &1.03 &1.09  \\
 FWA(xc)   &-0.0341  &-0.0344 &0.81 &1.81   \\
 FWA       &-0.0057  &-0.0058 &1.49 &0.96   \\
 SC      &-0.0026  &-0.0022 &15.29 &23.91  \\
 total     &-0.0083  &-0.0080 &3.76 &8.88  \\
\end{tabular}
\end{table}

\mediumtext
\begin{table}
\caption{Comparison between all-electron (AE) and pseudopotential
(PSP) dipole and quadrupole polarizability, in atomic units
($e^2=2$), for selected ions.
\label{sondrio}}
\begin{tabular} {lrrcrrc}
   & \multicolumn{3}{c} { Frozen-Wave Approximation }
   & \multicolumn{3}{c} { Self-Consistent Results} \\ \hline
   &   AE    & PSP   &error (\%)&   AE   &  PSP  &error (\%) \\
     \hline Dipole \\
 K$^+$
      &   8.89 &   $-$  &  $-$     &  5.74  &  $-$  &  $-$    \\
 K$^{+0.3}$ ($s^{0.7}$)
      & 181.94 & 175.76 &  3.4     & 165.89 & 164.83 &  0.6   \\
 C$^+$ ($s^2p^1$)
      &  10.04 &  10.16 &  1.2     &  6.04  &   6.04 &  0.0   \\
 Ar$^{+0.5}$ ($s^2p^{5.5}$)
      &  13.82 &  13.93 &  0.8     &  9.01  &   9.02 &  0.2   \\
 Si$^+$ ($s^2p^1$)
      &  31.61 &  31.82 &  0.7     &  19.11 &  19.19 &  0.4   \\
 Ga$^{+0.5}$ ($s^{2.0}p^{0.5}$)
      &  46.97 &  45.12 &  3.9     &  29.60 &  29.97 &  1.2   \\
 Ge$^+$ ($s^2p^1$)
      &  30.38 &  29.60 & 2.6      &  18.93 &  19.12 &  0.9   \\
 Ti$^+$ ($s^2p^0d^1$)
      &  90.50 &  90.61 &  0.1     &  47.98 &  48.26 &  0.6   \\
 \hline
 Quadrupole\\
 K$^+$
      &  18.7  &   $-$  &  $-$     & 18.2   &  $-$  &  $-$    \\
 K$^{+0.3}$ ($s^{0.7}$)
      &  2530  &  2521  &  0.3     & 3044   & 3029   &  0.5   \\
 C$^+$ ($s^2p^1$)
      &  16.4  &  16.4  &  0.0     & 16.8   &  16.8  &  0.0   \\
 Ar$^{+0.5}$ ($s^2p^{5.5}$)
      &  38.2  &  38.2  &  0.0     & 37.8   &  37.8  &  0.0   \\
 Si$^+$ ($s^2p^1$)
      &  107.5 &  107.5 &  0.0     & 108.2  &  108.3 &  0.1   \\
 Ga$^{+0.5}$ ($s^{2.0}p^{0.5}$)
      &  220.1 &  218.8 &  0.6     & 228.5  &  228.1 &  0.2   \\
 Ge$^+$ ($s^2p^1$)
      &  107.4 &  106.7 &  0.7     & 108.9  &  108.5 &  0.3   \\
 Ti$^+$ ($s^2p^0d^1$)
      &  301.8 &  286.8 &  5.0     & 319.9  &  316.1 &  1.2   \\

\end{tabular}
\end{table}

\narrowtext

\begin{figure}
\caption{Magnitudes of some important hardness matrix elements for
Ar in configuration $3s^{1.2}3p^{5.7}3d^{0.1}$.  The hollow bars
represent the all-electron results while the dashed bars represent
pseudopotential results at $r_{cs}=0.9$ a.u., $r_{cp}=
r_{cd}=1.0$ a.u. Different contributions
are decomposed into different columns as, ``h'' (Hartree), ``xc''
(exchange-correlation), ``FWA'' (total hardness under frozen
wavefunction approximation), ``SC'' (self-consistent correction), and
``total'' hardness.  The $L>0$ hardness matrix elements are
rescaled by a factor of $2L+1$ to make them more clear.
\label{f_ar}}
\end{figure}

\begin{figure}
\caption{Hardness matrix elements for Si$^+$ in configuration
$s^2p^1d^0$.
Hollow bars represent AE results, while the dashed and shaded
bars represent PSP results at
$r_{cs}=1.1$ a.u., $r_{cp}=1.2$ a.u., $r_{cd}=0.8$ a.u., and
$r_{cs}=2.3$ a.u., $r_{cp}=2.4$ a.u., $r_{cd}=0.8$ a.u., respectively.
\label{f_si}}
\end{figure}

\begin{figure}
\caption{Hardness matrix elements for O in configuration
$s^{1.4}p^{3.5}d^{0.1}$.
Hollow bars represent AE results, while the dashed and shaded
bars represent PSP results at $r_{cs}=0.8$ a.u.,
$r_{cp}=0.9$ a.u., $r_{cd}=0.8$ a.u., and
$r_{cs}=2.9$ a.u., $r_{cp}=3.0$ a.u., $r_{cd}=0.8$ a.u.,
respectively.
\label{f_o}}
\end{figure}

\begin{figure}
\caption{Hardness matrix elements for K in configuration
$s^{0.25}p^{0.25}d^{0.25}$.
Hollow bars represent AE results, while the dashed and shaded
bars represent PSP results ($r_{cs} = 1.8$ a.u., $r_{cp} = 2.3$ a.u.,
$r_{cd} = 1.2$ a.u.)
without and with the LFC partial-core correction,
respectively.
\label{f_k}}
\end{figure}
\begin{figure}
\caption{Hardness matrix elements for Ti in configuration
$s^1p^{0.25}d^2$.
Hollow bars represent AE results, while the dashed and shaded
bars represent PSP results ($r_{cs} = 1.8$ a.u., $r_{cp} = 2.3$ a.u.,
$r_{cd} = 0.8$ a.u.)
without and with the LFC partial-core correction,
respectively.
\label{f_ti}}
\end{figure}

\begin{figure}
\caption{Hardness matrix elements for C in configuration
$2s^{1.9},2p^{1.4},3d^{0.3}$.
Hollow bars represent AE results, while the dashed and shaded
bars represent PSP results ($r_{cs} = r_{cp} = 0.6$ a.u.,
$r_{cd} = 0.8$ a.u.)
without and with the LFC partial-core correction,
respectively.
\label{f_c}}
\end{figure}

\begin{figure}
\caption{Hardness matrix elements for Ga in configuration
$s^2p^{0.5}d^0$.
Hollow bars represent AE results, while the dashed and shaded
bars represent PSP results ($r_{cs} = 0.8$ a.u.,
$r_{cs} = 0.9$ a.u., $r_{cd} = 1.2$ a.u.)
without and with the LFC partial-core correction,
respectively.
\label{f_ga}}
\end{figure}

\begin{figure}
\caption{Calculated average hardness error for Ar, Si, Ga, and K as
a function of core radius used in the PSP generation. The
configurations used to calculate the average hardness errors are Ar
($3s^{1.95}3p^{5.95}3d^{0.05}$), Si ($3s^{1.2}3p^{2.7}3d^{0.1}$),
Ga ($4s^{1.2}4p^{1.7}3d^{0.1}$), and K
($4s^{0.85}4p^{0.1}3d^{0.05}$).
\label{f_av1}}
\end{figure}

\begin{figure}
\caption{Calculated average hardness error for C and O as a
function of core radius used in the PSP generation. The
configurations used to calculate the average hardness errors are C
($2s^{1.9}2p^{2.0}3d^{0.1}$) and O ($2s^{1.5}2p^{4.0}3d^{0.5}$).
\label{f_av2}}
\end{figure}
\end{document}